\documentclass[a4paper,twocolumn,11pt,accepted=2024-12-30]{quantumarticle}
\pdfoutput=1
\usepackage{graphicx}  
\usepackage{dcolumn}   
\usepackage{bm}        
\usepackage[utf8]{inputenc}
\usepackage[english]{babel}
\usepackage[T1]{fontenc}
\usepackage{amsmath}
\usepackage{tikz}
\usepackage{lipsum}

\usepackage[utf8]{inputenc}
\usepackage[english]{babel}
\usepackage[T1]{fontenc}
\usepackage{amsmath}
\usepackage{fixltx2e}
\usepackage[numbers,sort&compress]{natbib}

\usepackage{amsmath}
\usepackage[colorlinks=true, allcolors=blue]{hyperref}
 \usepackage{physics}

\begin{document}

\title{Unlocking Heisenberg Sensitivity with Sequential Weak Measurement Preparation}
\author{Trinidad B. Lanta{\~n}o}
\author{Dayou Yang}
\author{K. M. R. Audenaert}
\author{S. F. Huelga}
\author{M. B. Plenio}
\affiliation{Institut für Theoretische Physik, Albert-Einstein-Allee 11, Universität Ulm 89069, Germany}

\begin{abstract}
We propose a state preparation protocol based on sequential measurements of a central spin coupled with a spin ensemble, and investigate the usefulness of the generated multi-spin states for quantum enhanced metrology. Our protocol is shown to generate highly entangled spin states, devoid of the necessity for non-linear spin interactions. The metrological sensitivity of the resulting state surpasses the standard quantum limit, reaching the Heisenberg limit under symmetric coupling strength conditions. We also explore asymmetric coupling strengths, identifying specific preparation windows in time for optimal sensitivity. Our findings introduce a novel method for generating large-scale, non-classical, entangled states, enabling quantum-enhanced metrology within current experimental capabilities.
\end{abstract}

\maketitle

\begin{figure}[t!]
\centering{}\includegraphics[width=7.5cm]{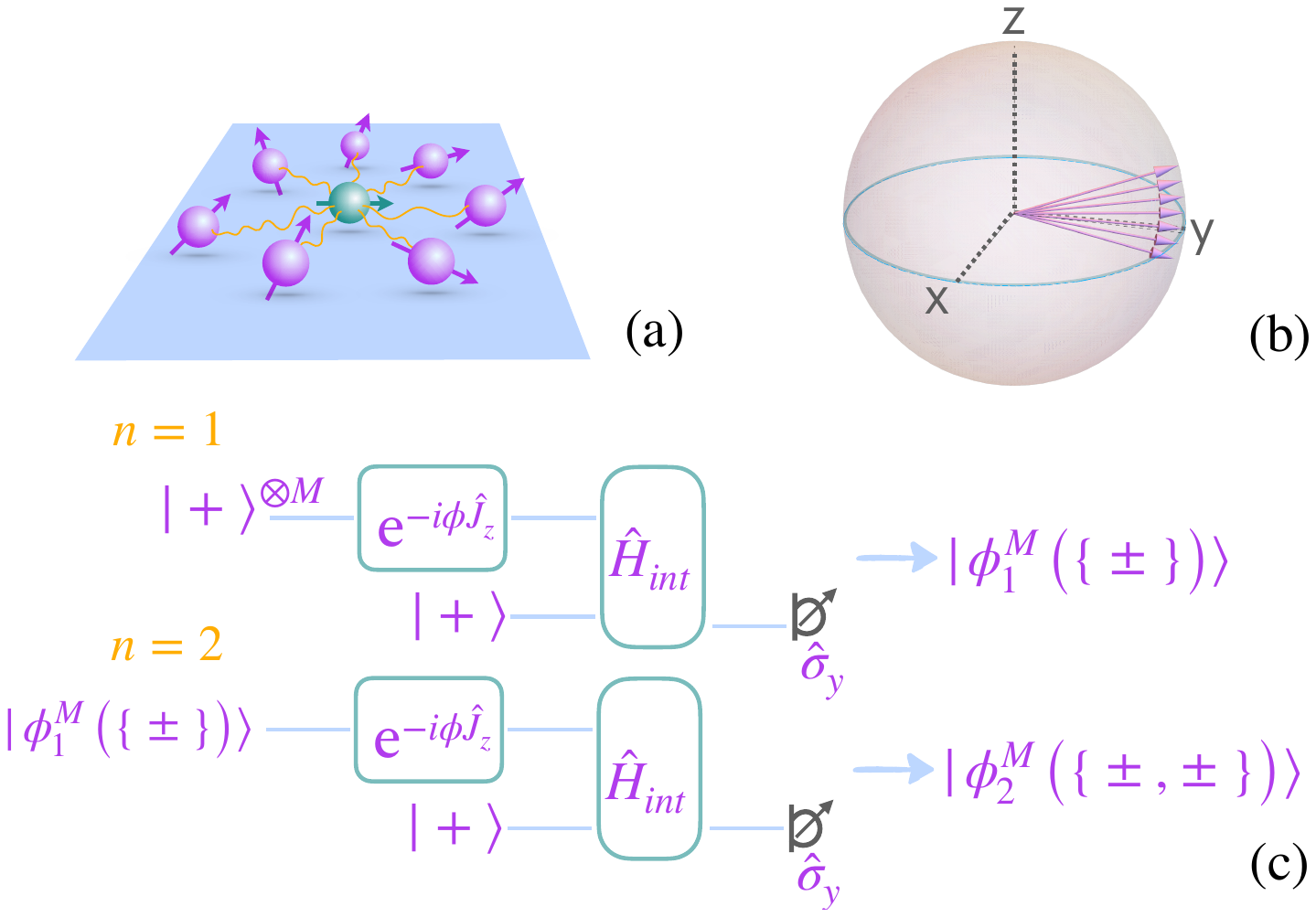}
\caption{\textbf{(a)} A central spin interacts with $M$ surrounding spins. We study the metrological usefulness of the system of the surrounding spins both, when the interaction coupling is symmetric and when it is non-symmetric.  \textbf{(b)} Bloch sphere representation of the spin coherent states that, in superposition, form the final state in the presence of a symmetric coupling. \textbf{(c)} Effective circuit of two successive runs in our proposed protocol for an M-particle entangled state.}
\label{fig:1}
\end{figure}

\emph{Introduction}---
Quantum metrology offers a unique opportunity to explore the quantum realm while making tangible contributions to technological innovations. 
Making use of the quantum mechanical properties such as optical \cite{Caves} and spin \cite{spinsqueez1,wineland1994squeezed} squeezing, multipartite entanglement \cite{huelgapartiallyent}, static phase transitions \cite{zanardi2008quantum, tsang2013quantum, chu2021dynamic} or continuously observed driven-dissipative phase transitions \cite{ilias2022criticality,yang2023efficient} 
we can extend the boundaries imposed by the standard quantum limit of classical statistics \cite{HaaseHuelga, Maccone}.\\
However, generating entanglement in multipatite many-particle systems is a highly non-trivial task experimentally. The process of entanglement generation necessitates controlled interactions among numerous particles, whose difficulty is typically increasing rapidly with the number of particles and the complexity of the targeted entangled state. One of the strategies for generating many-body entanglement is the so-called one-axis twisting (OAT). This method utilises a non-linear unitary transformation of the form $U(t)=\text{e}^{-it\xi J_i^2}$, with $J_i$ the spin operator of the collective system, which is applied to spin coherent states (SCS), which possess no entanglement, to let them evolve into spin squeezed states, \emph{i.e.}, entangled atomic states with reduced variance in one of the angular momentum components
 \cite{spinsqueez1, Wineland}. Such states can be realized, for example, using a two-mode Bose–Einstein
condensate \cite{BECsSqueezSorensen, Riedel2010,PhysRevLett.87.170402, spinsqueez}. 
Another notable family of entangled states in spin systems are \emph{spin cat states} \cite{spincatstates,catstates_N00N,spincatstates2}, namely macroscopic superpositions of two angular momentum states with different Bloch vectors \cite{catstates_inter}. These states are symmetric, in the sense that they are invariant under particle permutations, and thus can be described within the subspace of fixed total angular momentum. States with this symmetry were shown to be optimal for phase estimation in the presence of decoherence \cite{huelgapartiallyent,catstates_dissip}. Conventional approaches to generate spin cat states are adiabatic evolution in atomic Bose-Einstein condensates \cite{adiabatevol1, adiabatevol2, adiabatevol3} - which face the hurdle of limited coherence time and is thus far limited to a small number of particles \cite{adiabatevol1} - and twist-and-turn dynamics - where the OAT evolution is implemented along with additional rotations designed via machine optimization \cite{twistandturn}.\\ 
Here, we propose and investigate an alternative approach to the preparation of macroscopic superpositions of SCS. Inspired by the concept of quantum-hybrid sensors, \emph{i.e.}, devices that combine a central spin coupled to an assembly of spins, commonly utilized for sensing the precession angle of nuclear spins \cite{Cai2014,hybrid2,hybrid1,PhysRevA.101.032347, singlesiteaddress, diamondbiology}, we propose a protocol which performs repeated weak measurements \cite{datta, PhysRevLett.123.050603} of these nuclei via a central spin that is interacting coherently with each surrounding spin, thus enabling the preparation of a \emph{random} multipartite entangled state of the nuclei (Fig.\ref{fig:1} (a, b)). Note that the non-linear evolution such as the previously mentioned OAT dynamics has been replaced by a measurement based approach supplemented by linear dynamics only.
Systems consistent with this conceptual set-up have been realized experimentally in high resolution spectroscopy, where the electron spin of a Nitrogen vacancy (NV) center monitors the Larmor precession of surrounding $^{13}$C nuclear spins \cite{highresuspect, trackingprecess, nonlinspect,  schwartz2019, doi:10.1126/science.aam5532, boss2017quantum}.\\
Randomly prepared correlated multi-particle states via measurements has spurred intense theoretical and experimental  interests in recent years, in particular for understanding quantum chaos and thermalization~\cite{Choi2023,PhysRevLett.128.060601,PRXQuantum.4.010311}. On a purely theoretical level, generic random symmetric spin states have been shown to be metrologically useful~\cite{PhysRevX.6.041044}. Our work complements these efforts by establish a concrete experimental protocol for random state preparation for quantum-enhanced metrology, covering both symmetric and non-symmetric scenarios.
 
Our key findings are as follows: First, when the interaction between the central spin and $M$ surrounding spins is symmetric, our protocol prepares a superposition of states of the form $\ket{\theta, \varphi}^M=\left(\cos \frac{\theta}{2} \ket{0}+ \text{e}^{i \varphi} \sin \frac{\theta}{2} \ket{1}\right)^{\otimes M}$, where $\ket{\theta, \varphi}^M$ is a SCS. We refer to this family of states as \textit{multi-cat states}, because it includes spin cat states but generally contains states with more than two terms in the superposition. These states are invariant under particle exchange and, therefore, can be described within the fully symmetric Hilbert subspace of fixed total angular momentum $J=\frac{M}{2}$. Such pure symmetric states of qubits generically exhibit sub-shot noise sensitivity \cite{PhysRevA.82.012337, PhysRevX.6.041044}. Second and remarkably, we go beyond the symmetric paradigm and examine the achievable sensitivities within the more realistic scenario of non-symmetric couplings. In this case, the state generated by the protocol no longer belongs to the fully symmetric subspace. While previous findings suggested that, typically, random states outside the fully symmetric subspace are not useful for metrology \cite{PhysRevX.6.041044}, we discover that, for a finite preparation time and with all spins initialized in a SCS, the state gradually leaks out of the fully symmetric subspace. This allows for the state to maintain Heisenberg-limited (HL) precision for a certain time, eventually becoming standard quantum limited (SQL) for longer times. We provide time scales indicating when this transition occurs, showing an inverse relation to the amount of inhomogeneities present in the couplings.

\emph{Protocol}---Consider an ensemble of $M$ non-interacting spins initialized in a product state $\vert + \rangle^{\otimes M}$, where $\ket{+}=(\ket{0}+\ket{1})/\sqrt{2}$. Subsequently, the spin ensemble undergoes $n$ repeated evolution cycles, as shown in Fig.~\ref{fig:1}(b), with each cycle consisting of three successive steps: (i) The $M$ spins freely precess around the $z$ axis by an angle $\phi$, described by the unitary operation $U_\phi = \text{exp}\left[{-i\frac{\phi}{2} \sum_{k=1}^M \sigma_z^{(k)}}\right]$. (ii) An additional spin, which we refer as to central spin, is initialized in the state $\ket{+}$. Under the application of a XY-sequence, the central spin then interacts with the $M$ spins via the Hamiltonian 
\begin{equation}
\label{eq:Hint}
H_{\rm int}=\sigma_z \otimes \sum_{k=1}^M \alpha_k\sigma_x^{(k)}
\end{equation}
for a duration $T$, where $\sigma_z$ acts on the central spin and the $\sigma_x^{(k)}$ act on the k-th spin (see \cite{PhysRevA.101.032347} and Appendix \ref{AppendixA} for details of the derivation). (iii) Perform a projective measurement on the central spin along the $\sigma_y$ basis, providing the binary measurement outcome $s=\pm 1$, corresponding to the central spin being up (down). These steps (i-iii) can be naturally realized with interacting spin systems in various quantum-optical and solid-state platforms, such as a $^{13}{\rm C}$ nuclear spin ensemble coupled with the electron spin of a nearby NV center in diamond \cite{highresuspect, PhysRevA.101.032347}. In this case, $\alpha_k$ in Eq. (\ref{eq:Hint})
corresponds to the perpendicular coupling of the nuclear spins with the NV center, the reader can refer to \cite{trackingprecess} for an experimental realization of Hamiltonian (\ref{eq:Hint}). Note that while the nuclear-NV coupling depends on the nuclear position with respect to the NV-center, it can neverthless be reduced to Eq. (\ref{eq:Hint}) via a local unitary transformation for each nuclear spin (see Appendix \ref{AppendixA} for details).\\
The protocol is performed sequentially $n$ times and therefore defines a discrete \emph{quantum trajectory} consisting of a
series of measurement records $\mathbf{S}=\{s_1,s_2,\dots,s_n\}$ and the associated conditional evolution of the nuclear spin ensemble.\\
Note that each run of the protocol is equivalent to applying the operator $\hat{O}_{s_j}=\frac{1}{2}\left(\text{e}^{-\sum_{k=1}^M \alpha_k\sigma_x^{(k)}} + i s_j \text{e}^{\sum_{k=1}^M \alpha_k\sigma_x^{(k)}}\right)\text{e}^{-i\frac{\phi}{2} \sum_{k=1}^M \sigma_z^{(k)}}$ onto the SCS $\ket{\theta=\frac{\pi}{2}, \varphi=0}$. Therefore, when $\phi=0$, no entanglement is created as the SCS is an eigenstate of $\hat{O}_{s_j}$, conversely, if $\phi=\frac{\pi}{2}$, we expect that the rotation angle around the $x$ axis of each spin will be maximized, causing the state to spread further from the equator in the Bloch sphere representation. Given a SCS, each cycle of the conditional evolution maps the original SCS into a superposition of two SCSs pointing in different directions, cf. Fig.~\ref{fig:1}(c). The conditional state after $n$ measurements is therefore a superposition of $2^n$ SCSs, which we refer to as a spin \emph{multi-cat state},
\begin{align} \label{eq:sym_state}
    \ket{\phi (\mathbf{S})} = \sum_{i=1}^{2^n} w_i (\mathbf{S}) \ket{\psi_i}^{\otimes M}
\end{align}
where $w_i$ are complex weights defined in eq. (\ref{eq:complexweights_def}) that depend on the string of outcomes, and $\ket{\psi_i}$ is a SCS in the $i$-th direction (see the Appendix \ref{app:stateexpression} for details).

Below we investigate the capacity of the generated multi-cat states usefulness to surpass the SQL in quantum metrology. The relevant figure of merit is the Quantum Fisher Information (QFI) \cite{Tóth_2014,Maccone}
\begin{align}\label{eq:FQ_def}
    F_Q \left[ \ket{\phi (\mathbf{S})}; \hat{J}_{\hat{n}} \right]= 4 \Delta^2 \hat{J}_{\hat{n}},
\end{align}
which measures the sensitivity of $\ket{\phi (\mathbf{S})}$ to a rotation $\ket{\phi (\mathbf{S})} \rightarrow \text{e}^{-i \theta \hat{J}_{\hat{n}}} \ket{\phi (\mathbf{S})} \equiv \ket{\phi_\theta (\mathbf{S})} $, with $\hat{J}_{\hat{n}} = \frac{1}{2} \sum_{i=1}^M \hat{\sigma}_{\hat{n}}^{(i)}$. As shown in Appendix \ref{app:vanishingJz}, state (\ref{eq:sym_state}) exhibits a mirroring symmetry with respect to the XY plane, which implies $\expval{\hat{J}_z}=0$.  It is for this reason that later on, we compute the QFI for $\hat{n}=z$.  Typically, the rotation angle corresponds to $\theta = B_0 t$, where $B_0$ is the magnitude of a magnetic field the system is subjected to for a time duration $t$. By measuring many copies of the state $\ket{\phi_\theta (\mathbf{S})}$,  a probability distribution is obtained allowing the estimation of $B_0$.
According to the quantum Cr\'amer-Rao inequality~\cite{cramer}, the QFI sets an upper limit to the estimation precision achieved by any measurement strategy of the quantum state $\ket{\phi (\mathbf{S})} $,
$ \Delta^2 \Tilde{\theta} \geq ( F_{Q})^{-1}$. Moreover, the QFI Eq.~\eqref{eq:FQ_def} is also a witness for multipartite entanglement \cite{pezzeQFIEntangl}: $\ket{\phi (\mathbf{S})}$ is at least $\lfloor F_{Q}/M\rfloor$ partite entangled, with $\lfloor x\rfloor$ indicating the largest integer not greater than $x$. A state with vanishing entanglement, a separable state, satisfies the so-called \textit{standard quantum limit} of precision $F_Q \leq M$ \cite{PhysRevLett.102.100401}. 
On the other hand, GHZ states \cite{Greenberger1989} saturate the \emph{Heisenberg limit} $F_Q = M^2$. Interestingly, a state with an amount of entanglement that vanishes at a suitably chosen rate with increasing number of particles, can attain any precision scaling arbitrarily close
to the Heisenberg Limit \cite{asymptent}.

As a measure of the overall metrological usefulness of the ensemble of stochastic states $\ket{\phi({\mathbf{S} })}$, we focus in the following on the \emph{average QFI} over all possible measurement outcomes $\mathbf{S}$,
\begin{equation}\label{eq:Avgqfi}
\overline{F}_{Q}^{(n)}:=\sum_{\mathbf{S}}P(\mathbf{S})F_Q \left[ \ket{\phi (\mathbf{S})}; \hat{J}_{\hat{n}} \right],
\end{equation}
where $P(\mathbf{S})= \braket{\phi (\mathbf{S})}{\phi (\mathbf{S})}$ is the probability of recording the outcome $\mathbf{S}$. We are primarily interested in analyzing the scaling behavior of the average QFI with respect to the number of spins $M$ and the number of measurement cycles $n$. The QFI in Eq. (\ref{eq:Avgqfi}) represents the statistical average of the optimal precision achievable by a randomly generated state, and this precision can be attained by selecting an optimal measurement that adapts to each individual trajectory.
\begin{figure}[t!]
\includegraphics[width=7cm]{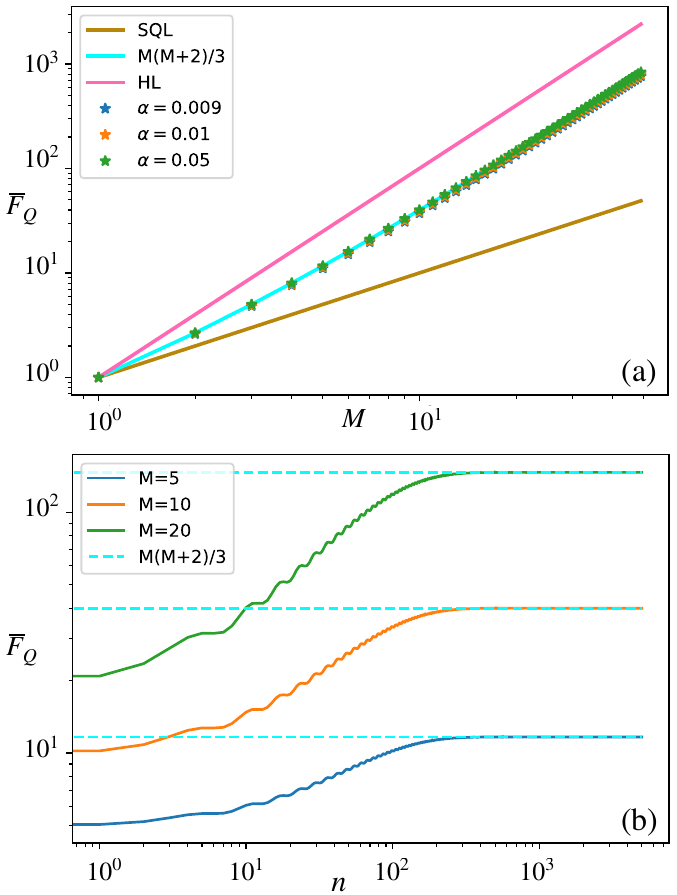}
\caption{\label{fig:QFI_symm} QFI averaged over all measurement outcomes after $n=5000$ repetitions of the protocol: \textbf{
(a)} As a function of the number of particles $M$, the pink (blue) solid lines are the Heisenberg (standard quantum) limits $\overline{F}_Q^{(n)}=M^2 (M)$, the cyan line is our result for large $n$ (eq. (\ref{eq:QFIsym_longtime})) and the stars are obtained from eq. (\ref{eq:QFI_symmm}) at a fixed $n$. Three couplings strengths $\alpha$ are considered in Eq. (\ref{eq:Hint}). \textbf{(b)} For a coupling strength $\alpha=0.05$ in Eq. (\ref{eq:Hint}), the dashed lines represent Eq. (\ref{eq:QFIsym_longtime}) and the colored solid lines are obtained from Eq. (\ref{eq:QFI_symmm}) at a fixed number of particles $M$. In both figures we used a precession phase of $\phi = 0.5$ per cycle (as described in the section \textit{protocol}).}
\end{figure}

\emph{Symmetric coupling}---The operation principle of our state-generation protocol is most transparent in the case where all nuclear spins couple with the electron spin with identical interaction strength; i.e., $\alpha_k=\alpha$ in Eq.~\eqref{eq:Hint}.
Noting that the conditional state Eq.  (\ref{eq:sym_state}) is invariant under the transformation $Z\leftrightarrow -Z$, we obtain $\expval{J_z} = 0$ (Appendix \ref{app:vanishingJz}). Utilizing this feature, we can rigorously show (see Appendix \ref{app:AvgQFI} for a detailed derivation) that the average QFI follows a quadratic scaling with respect to the number of particles
\begin{align} \label{eq:QFI_symmm}
    \overline{F}_{Q}^{(n)}= M + M(M-1) \overline{H}_S(n),
\end{align}
with $\overline{H}_S(n) := \Tr \overline{\varrho}_n \sigma_z^{\otimes 2}$ and $\overline{\varrho}_n$ is a two particle operator given by Eq. (\ref{eq:rho2partavg_sym}). It is clear from its definition that $\overline{H}_S(n)$ does not depend on the number of particles $M$, and moreover, as shown in Appendix \ref{app: H2avg}, it is a positive, non-vanishing quantity, that does not decrease with $n$.  Furthermore in Appendix \ref{app:AvgQFI}, we derive an analytical expression for the long-time $\overline{F}_{Q}^{(n)}$, and obtain
\begin{align}\label{eq:QFIsym_longtime}
    \overline{F}_Q^{(\infty)} = \frac{1}{3} M(M+2).
\end{align}
At the end of the aforesaid Appendix,  we show that Eq. (\ref{eq:QFIsym_longtime}) together with Eq. (\ref{eq:QFI_symmm}) imply $\overline{H}_S(n\rightarrow \infty)=\frac{1}{3}$. 
 Our result shows that $\overline{F}_{Q}^{(n)}$ 
 converges towards Heisenberg scaling with respect to the number of spins $M$ cf. Fig.(\ref{fig:QFI_symm}). This confirms that our protocol can generate multipartite entanglement extending across the entire system. Another remarkable feature is that, since $\overline{H}_S(n)$ is independent of the number of spins $M$, the number of repetitions $n$ required to saturate the asymptotic value only depends on the parameters $\alpha_k=\alpha$ and $\phi$ of the protocol (see section \textit{protocol}); this is demonstrated in Fig. \ref{fig:QFI_symm}(b). As such, our protocol enables \emph{scalable} generation of metrologically useful, multipartite entangled resource states. The values for coupling $\alpha$ in  Fig. (\ref{fig:QFI_symm}) were taken from an experimental realization of Hamiltonian (\ref{eq:Hint}) with NMR spectroscopy in diamond \cite{trackingprecess}.

\begin{figure}[t!]
  \includegraphics[width=7cm]{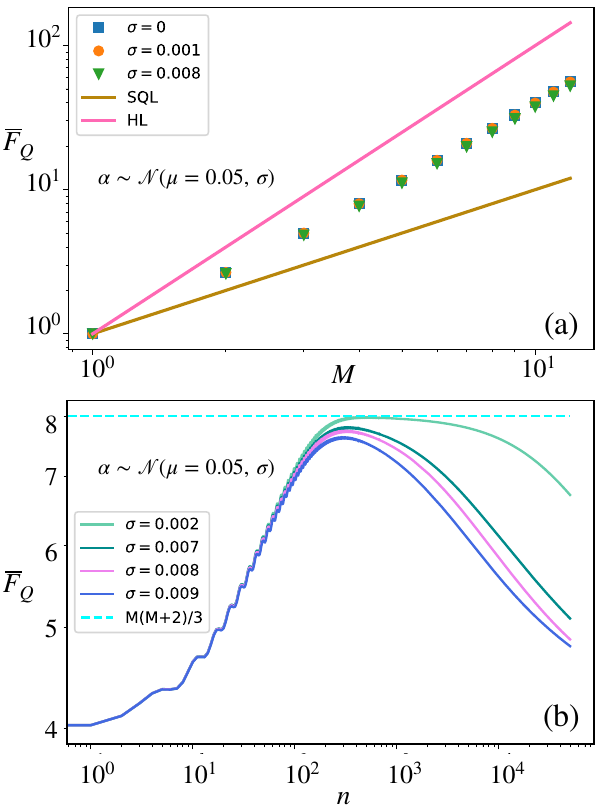}
  \centering
  \caption{QFI averaged over all measurement outcomes (eqs. (\ref{eq:avgQFI_nonsym}) and (\ref{eq:numQFI_nonsym})): \textbf{(a)} For $n=500$ repetitions of the cycle and $\phi=0.5$ in (\ref{eq:T^k_nonsymmetric}), two normal distributions centered at $\overline{\alpha} = 0.05$ with standard deviations $\sigma=0.001$ and $\sigma=0.008$ along with the symmetric coupling case $\sigma=0$ are shown. \textbf{(b)} As a function of the number of cycles, with $M=4$ and $\phi = 0.5$. For $\alpha_k$ we consider four normal distributions centered at $\overline{\alpha} = 0.05$ with $\sigma=0.002$, $\sigma=0.007$, $\sigma=0.008$, and $\sigma=0.009$.}
  \label{fig: QFI_nonsym}
\end{figure}

\emph{Non-Symmetric coupling}---The actual implementation of our protocol in realistic experimental platforms will typically feature non-symmetric couplings, i.e., disorder in the coupling strength $\alpha_k$ in Eq.~\eqref{eq:Hint}, due to either the position disorder of the spins or the angular inhomogeneity of the interaction (e.g., dipolar coupling). As we show below, the proposed state-generation protocol can still generate robust multipartite entanglement in the presence of non-symmetric coupling. Furthermore, such entanglement generation process demonstrates interesting transient dynamics that is absent in the case of symmetric coupling.\\ 
To this end, we study again the conditional multi-cat state $\ket{\phi({\mathbf{S}})}$ associated with a specific measurement record $\mathbf{S}$, now with a non-symmetric coupling $\alpha_k\neq\alpha_j$ in Eq. (\ref{eq:Hint}). The conditional state can be written in this case as (Appendix \ref{app:stateexpression})
\begin{align} \label{eq:nonsym_state}
    \ket{\phi (\mathbf{S})} = \sum_{i=1}^{2^n} w_i (\mathbf{S}) \bigotimes_{k=1}^M \ket{\psi_i^k}
\end{align}
where the $w_i$ are again  the complex weights in Eq. (\ref{eq:complexweights_def}) and $\ket{\psi_i^k}$ denote a single-spin SCS in the $i$-th direction associated with the $k$-th spin.
Using the parametrization in Eq.~\eqref{eq:nonsym_state}, we obtain in Appendix \ref{app:AvgQFI} an analytical expression for the QFI averaged over all possible measurement outcomes as
\begin{align} \label{eq:avgQFI_nonsym}
    \overline{F}_{Q}^{(n)}  = M + 2\overline{H}_{NS}(M, n),
\end{align}
where $\overline{H}_{NS}(M, n)$ is given by eqs. (\ref{eq:h2_def_nonsym})(\ref{eq:hNS_nonsym}) and it is a positive number not greater than ${M(M-1)}/2$ (shown in Appendix \ref{app: H2avg}). The factor $\overline{H}_{NS}(M, n)$ depends nontrivially on the number of repetitions $n$: it maintains a quadratic scaling with respect to the system size $M$, akin to the symmetric case [cf. Eq.~\eqref{eq:QFI_symmm}], for a transient time window $t\leq t_c$; and vanishes asymptotically for $t\gg t_c$. Therefore, the average QFI for non-symmetric coupling features superlinear precision scaling with respect to the system size in the transient regime, and a linear scaling in the long time limit, where
\begin{align} \label{eq:QFI_nonsym_asymp} 
\overline{F}_Q^{(\infty)} = \frac{1}{2^{M-2}} \Tr J_z^2 = M.
\end{align}
Such transient behavior is illustrated in Fig. (\ref{fig: QFI_nonsym}), where we considered the set of couplings $\{\alpha_k \}$ to be normally distributed with the same mean. We compare the metrological performance (as measured by Eq. (\ref{eq:avgQFI_nonsym})) between distributions with different standard deviations, which we refer to as \emph{disorder strength}.\\
In Figure (\ref{fig: QFI_nonsym}.b) we observe that as the disorder strength, i.e. $\sigma$, increases, the precision - as given by $\overline{F}_Q$ - decays faster from its maximum. Conversely, a smaller disorder strength allows for an enhanced precision sustained over $n$ (although we know from Eq. (\ref{eq:QFI_nonsym_asymp}) that eventually $\overline{F}_Q$ is going to drop to the linear scaling). In order to quantify these time scales, we can consider the continuous-time limit of our state-generation protocol, which allows us to describe the evolution of the average QFI in terms of a master equation, as discussed in the next section.

\emph{Master equation description}---To further elucidate the finite-time dynamics of our state-generation protocol, let us consider the limit of frequent repetition, i.e., $dt\equiv T/n\to0$ and correspondingly $\phi, \alpha \ll 1$. In this regime, we can expand the evolution of the conditional state $\ket{\phi({\mathbf{S}})}$ [see Eq. (\ref{eq:symstate_kraus}) for details] in $\phi,\alpha$ up to first and second order respectively.

Since $\expval{J_z}=0$  for our state (Appendix \ref{app:vanishingJz}), we can infer the average QFI in Eq. (\ref{eq:Avgqfi}) from the normalized average state (details in Appendix \ref{app:AvgQFI})
\begin{align}
     \overline{\rho}_n := \sum_{\textbf{S}} \ketbra{ \phi_n(\mathbf{S})}{\phi_n(\mathbf{S})}.
\end{align}

In the case of symmetric coupling, we show in Appendix \ref{app:MEderivation} that the time evolution of $\overline{\rho}(t) :=\bar{\rho}^{\rm S}$ can be expressed in the Lindblad form, namely
\begin{align}\label{eq:master_eq_sym}
    \frac{ d\bar{\rho}^{\rm S} }{dt}= -i \left[\tilde{\phi} J_z, \; \overline{\rho}^{\rm S}  \right] + \gamma \mathcal{D} \left[J_x \right]\overline{\rho}^{\rm S}
\end{align}
where $\tilde{\phi}:= \phi/dt$, $\gamma:=4\alpha^2/dt$ and the Lindblad dissipator is defined via $\mathcal{D}\left[L\right]\rho := L \rho L^\dagger - \frac{1}{2} \{L^\dagger L, \rho \}$. 
Analogously, in the presence of non-symmetric couplings, we expand the corresponding evolution of $\ket{\phi({\rm \textbf{S}})}$ [see Eq.~\eqref{eq:state2}] and write  $\alpha_k = \alpha + \delta_k$ to obtain a master equation for the average state
\begin{align}
\label{eq:MEdisorder} 
 \frac{d\overline{\rho}^{\rm NS} }{dt}= &-i \left[\tilde{\phi} {J}_z, \; \overline{\rho}^{\rm NS}  \right] + \gamma \mathcal{D}\left[J_z\right] \overline{\rho}^{\rm NS} \\ \nonumber & + \sum_{k=1}^M  \Gamma_k \mathcal{D}\left[ \sigma_x^{(k)}\right]\overline{\rho}^{\rm NS},
\end{align}
where $\Gamma_k=4 \overline{\delta_k^2}/dt$.\\ 
Equation~\eqref{eq:MEdisorder} allows us to distinguish two characteristic time scales. First, a short time scale $t_s \simeq 1/\gamma=\text{d}t/4\alpha^2$, corresponding to the repetition number $n_s \simeq 1/{\alpha^2}$, where the right-hand-side (RHS) of Eq.~\eqref{eq:MEdisorder} is dominated by the first two terms; i.e., identical to Eq.~\eqref{eq:master_eq_sym} for the case of homogeneous coupling. As a result, the average QFI obeys the HL with respect to the system size $M$. Second, a long time scale $t_l \simeq 1/\Gamma_k=\text{d}t/\overline{\delta_k^2}$,corresponding to the repetition number $n_l  \simeq 1/\overline{\delta_k^2} $, where the RHS of Eq.~\eqref{eq:MEdisorder} is dominated by the disorder. In this regime, the advantageous scaling with respect to $M$ drops steeply to the SQL. In particular, when we consider a normal distribution for the couplings and quantify the disorder strength with the standard deviation $\sigma$ as in Fig. (\ref{fig: QFI_nonsym}), Eq.~\eqref{eq:MEdisorder} predicts that the optimal scaling is going to decrease dramatically at  $n_l  \simeq 1/ \sigma^2$. Therefore, we have verified and quantified what we observed numerically: for disordered coupling strengths, there is a finite transient regime where the QFI of the generated state is Heisenberg-limited, and the duration of this transient window is inversely proportional to the disorder strength. Such quantitative prediction provides a guideline for the practical implementation of our state-generation scheme in realistic experimental platforms in the presence of disorder.

\emph{Conclusion}---
Within the established framework of quantum-hybrid sensors, where a central spin couples to a large number of auxiliary spins, we have devised a strategy for preparing large-scale auxiliary spin-entangled states. This strategy is based on repeated measurements of the central spin, and we examine the potential of these states for metrology. Our work identifies the conditions under which the metrological sensitivities of these states exceed the standard quantum limit, thus indicating their application potential.

This suggests several avenues for future research. On the one hand, it will be intriguing to explore if and to what extent the utility of the macroscopic superpositions of spin-coherent states, prepared in our scheme, may be linked to the negativity of their Wigner representation. This connection is based on earlier observations of similar relationships in continuous variable systems. More generally, this also prompts the question of whether and to what extent these states are more resourceful than spin-squeezed states.

On the other hand, given the application potential of these states, it is pertinent to study specific physical implementations of this scheme. This spans a range from color centers in diamond to trapped ions, all while considering experimentally realistic parameters and imperfections.\\
\emph{Acknowledgments}--- This work was supported by the ERC synergy grant HyperQ (grant no. 856432), the BMBF project PhoQuant (grant no. 13N16110) and the EU projects QuMicro (grant no. 01046911) and C-Quens (grant no. 101135359).

\appendix
\section{A specific physical model: Nitrogen vacancy (NV) sensor} \label{AppendixA}
The Nitrogen Vacancy Center, or NV center, is exclusively located within diamond crystals, where carbon atoms serve as a nuclear spin bath. This unique characteristic, and the possibility for single site addressability \cite{singlesiteaddress}, renders it an ideal configuration for investigating the central spin model.\\
Let us start with the Hamiltonian of an NV center coupled to $M$ nuclear spins \cite{PhysRevA.92.042304} in the interaction picture
\begin{align}
H =& \frac{\Omega (t)}{2} \sigma_x + \sum_{m=1}^M \Bigg(\frac{\omega_L}{2} \sigma_z^{(m)} + \sigma_z \otimes \Vec{A}_{\perp}^{(m)} \cdot \Vec{\sigma}^{m}\Bigg)
\end{align}
where the first term is the contribution from the microwave driving and $\Omega (t)$ is the associated Rabi frequency. The second term is the energy contribution from the nuclear spins, where $\omega_L$ is their Larmor frequency. The last term describes the interaction between the NV and nuclear spins, where  $\Vec{\sigma}^{m}$ is the spin vector of the $m$-th nuclear spin and $\Vec{A}_{\perp}^{(m)}$ is the perpendicular part of the hyperfine vector, given by
\begin{align}
\Vec{A}_{\perp}^{(m)} = -\frac{3 \mu_0 \gamma_e \gamma_n}{4\pi \abs{\Vec{r}_m}^3} \frac{(\hat{z}\cdot \Vec{r}_m)}{\abs{\Vec{r_m}}^2}\left[(\hat{x}\cdot \Vec{r}_m)\hat{x} + (\hat{y}\cdot \Vec{r}_m)\hat{y} \right]
\end{align}
 where $\Vec{r}_m$ is the vector that connects the position of the NV-center with the $m$-th nuclear spin, $\gamma_e$($\gamma_n$) the electronic(nuclear) gyromagnetic ratio, and $\mu_0$ the vacuum magnetic permeability.
Now we use a dynamical decoupling (DD) sequence to create an effective $\sigma_z \otimes \sigma_x$ interaction \cite{PhysRevA.101.032347}. When the duration of the DD (e.g. XY-8) sequence fulfils $\tau = \frac{\pi}{\omega_L}$ the efective Hamiltonian can be expressed as
\begin{align} \label{eq:heff_app}
H_{\text{eff}} &= \sigma_z \otimes \frac{2}{\pi}\sum_{m=1}^M \Vec{A}_{\perp}^{(m)} \cdot \Vec{\sigma}^{m} \nonumber \\& \equiv \sigma_z \otimes \sum_{m=1}^M \beta_m \sigma_{\hat{n}_m}  
\end{align}
where $ \sigma_{\hat{n}_m}:=\sigma_x^{(m)}\cos \left(\phi_{0m}\right) + \sigma_y^{(m)}\sin \left(\phi_{0m}\right)$ with $\phi_0$ the corresponding angle between the $X$-$Y$ projection of $\Vec{r}_m$ and the $x$ axis, and $\beta_m := \frac{2}{\pi} \abs{\Vec{A}_{\perp}^{(m)}}$. 
By rotating the local coordinate of each nuclear spin such that  $ \sigma_{\hat{n}_m} = \sigma_{xm}$ where $\sigma_{xm}$ is the $\sigma_x$ operator of the $m$-th nuclear spin, we can reduce Eq. (\ref{eq:heff_app}) to Eq. (\ref{eq:Hint}) of the main text.\\
We can use the NV center to initialize the nuclear spins in $\ket{+}^{\otimes N}$ in the locally rotated basis as follows: Let us consider the symmetric case where all the $\beta_m$ are the same, i.e. $\beta_m=\beta$, which can be approximately achieved by arranging the nuclear spins in a cylindrically symmetric setting and placing the electron spin along the central axis. If we initialize the nuclear spins and the NV-center in the state $\ket{0}$, the interaction through Eq. (\ref{eq:Hint}) yields
\begin{align}
    e^{-i t H_{\text{int}}} \ket{0}\otimes \ket{0}^{\otimes M} =\ket{0} \otimes \left[  e^{-i t \beta \sum_{m=1}^M  \sigma_{xm}} \ket{0}^{\otimes M} \right]\, .
\end{align}
By letting the coupled system evolve for a time $t =\frac{\pi}{4\beta}$, followed by a precession around the $z$ axis, the state of the nuclear spins is mapped to $\ket{+}^{\otimes M}$ in the rotated basis.
Finally, we drive the state of the NV center to $\ket{+}$ to recover the initial configuration of the protocol in the main text.

\section{State generated by the protocol}\label{app:stateexpression}
\subsection{Symmetric coupling}
Each stage of the protocol acts on the state as a map 
\begin{align}
\hat{T}_{s_j} &=  \frac{1}{2}\left(  T_0^{\otimes M} + i s_j  T_1^{\otimes M} \right)\, ,
\end{align}
where we have defined the one-spin operators
\begin{align}
    T_0 &:= e^{-i\alpha \sigma_x} e^{-i\frac{\phi}{2} \sigma_z}\, , \nonumber \\
    T_1 &:= e^{i\alpha \sigma_x} e^{-i\frac{\phi}{2} \sigma_z}
\end{align}
and $s_j = \pm 1$ depending on the measurement outcome. The resulting (non-normalized) state, conditional to the set of outcomes $\mathbf{S} = (s_1, s_2... s_n )$, is
\begin{align}\label{eq:symstate_kraus}
 \ket{\phi_n (\mathbf{S})} = \frac{1}{2}\left(   T_0^{\otimes M} + i s_n  T_1^{\otimes M}\right) \ket{\phi_{n-1} (\mathbf{S})}
\end{align}
with probability
\begin{align}
    P(\mathbf{S}) = \braket{\phi_n (\mathbf{S})}{\phi_n (\mathbf{S})}.
\end{align}
Now, let us define
\begin{align}
    \ket{\psi_i} :=\left( \prod_{q=1}^n T_{i_q} \right)\ket{+}
\end{align}
and 
\begin{align} \label{eq:complexweights_def}
    w_i (\mathbf{S}) = \frac{1}{2^n} \prod_{q=1}^n (i s_q)^{i_q}\, ,
\end{align}
where $i = (i_1, i_2... i_n)$ is a binary string indicating which operator is acting on the state. With this we can write the complete state as
\begin{align}
    \ket{\phi (\mathbf{S})} = \sum_{i=1}^{2^n} w_i (\mathbf{S}) \ket{\psi_i}^{\otimes M}.
\end{align}
\subsection{Non-symmetric coupling}
The Kraus map performed on the state at each cycle of the protocol is now
\begin{align}\label{eq:nonsym_kraus}
\hat{T}_{s_j} =  \frac{1}{2}\left(  \bigotimes_{k=1}^M T_0^k + i s_j \bigotimes_{k=1}^M T_1^k \right)
\end{align}
with the one-spin operators
\begin{align}\label{eq:T^k_nonsymmetric}
    T_0^k &:= e^{-i\alpha_k \sigma_x^{(k)}} e^{-i\frac{\phi}{2} \sigma_z^{(k)}}\, ,\nonumber \\
    T_1^k &:= e^{i\alpha_k \sigma_x^{(k)}} e^{-i\frac{\phi}{2} \sigma_z^{(k)}}
\end{align}
and $s_j = \pm 1$. In a similar fashion to the symmetric case, the resulting (non-normalized) state, conditional to $\mathbf{S} = (s_1, s_2... s_n )$, is
\begin{align}\label{eq:state2}
 \ket{\phi_n (\mathbf{S})} = \frac{1}{2}\left(  \bigotimes_{k=1}^M T_0^k + i s_n \bigotimes_{k=1}^M T_1^k \right) \ket{\phi_{n-1} (\mathbf{S})}\, .
\end{align}
Moreover, let us define
\begin{align}\label{eq:nonsym_component}
    \ket{\psi_i^k} :=\left( \prod_{q=1}^n T_{i_q}^k \right)\ket{+}
\end{align}
where $i = (i_1, i_2... i_n)$ is again a binary string. This yields the desired representation of the state
\begin{align} 
    \ket{\phi (\mathbf{S})} = \sum_{i=1}^{2^n} w_i (\mathbf{S}) \bigotimes_{k=1}^M \ket{\psi_i^k}\, .
\end{align}
\section{Vanishing $\expval{J_z}$} \label{app:vanishingJz}
\subsection{Symmetric coupling} 
First, recall the definitions:
\begin{align*}
    T_0 &:= e^{-i\alpha \sigma_x} e^{-i\frac{\phi}{2} \sigma_z}\, , \nonumber \\
    T_1 &:= e^{i\alpha \sigma_x} e^{-i\frac{\phi}{2} \sigma_z}
\end{align*}
with which the Kraus map for the j-th run of the protocol is
\begin{align*}
    T_{s_j} = \frac{1}{2} \left(T_0^{\otimes M} + is_j T_1^{\otimes M} \right)\, .
\end{align*}
Now, consider the symmetry operation of \textit{mirroring} with respect to the $X-Y$ plane in the Bloch sphere representation. For a qubit vector the mirroring operation $\mathcal{M}$ would give
\begin{align*}
   \mathcal{M}\left(\begin{bmatrix} 
          \alpha \\
          \beta
         \end{bmatrix}\right)=
   \begin{bmatrix} 
          \beta^* \\
          \alpha^*
         \end{bmatrix}\, ,
  \end{align*}
thus leaving the $x,y$ coordinates unchanged while changing the sign of $z$. If $\ket{x}\in \mathcal{H}^{\otimes M}$ is a spin coherent state, then
  \begin{align*}
   \mathcal{M}\left(\ket{x}\right)= \sigma_x^{\otimes M} \ket{x^*}
  \end{align*}
where $\ket{x^*}$ means entry-wise complex conjugation. Also, note that
\begin{align*}
J_z \sigma_x^{\otimes M} &= -\sigma_x^{\otimes M} J_z\, , \\
J_x \sigma_x^{\otimes M} &= \sigma_x^{\otimes M} J_x
\end{align*}
where $J_i = \frac{1}{2} \sum_{m=1}^M \sigma_i^{(m)}$. Therefore
\begin{align}
    T_{s_j} \sigma_x^{\otimes M} &= \frac{1}{2} \left(T_0^{\otimes M} + is_j T_1^{\otimes M} \right)\sigma_x^{\otimes M}\nonumber \\
    &= \sigma_x^{\otimes M} i s_j  \left(-i s_j T_1^{\otimes M^*} +  T_0^{\otimes M^*} \right) \nonumber \\ &= i s_j \sigma_x^{\otimes M} T_s^*
\end{align}
hence, for any spin coherent state $\ket{x}$
\begin{align} \label{eq:mirrormap_sym}
    T_{s_j} \mathcal{M} \left( \ket{x} \right) = i s_j \mathcal{M} \left( T_{s_j} \ket{x} \right)\, .
\end{align}
Now, consider the expectation value of $J_z$ in some state $\ket{\psi'} = \mathcal{M} \left(\ket{\psi}\right)$, then 
\begin{align} \label{eq:proof_expvalzero}
    \bra{\psi'} J_z \ket{\psi'} =  \bra{\psi*}\sigma_x^{\otimes M} J_z \sigma_x^{\otimes M} \ket{\psi^*} = - \bra{\psi} J_z \ket{\psi}\, .
\end{align}
Therefore if the initial state is invariant under $\mathcal{M}$, as is $\ket{\phi_0} = \ket{+}^{\otimes M}$, $\expval{J_z}$ must vanish at every stage of the protocol.

\subsection{Non-symmetric coupling}
Here, the Kraus map changes to 
\begin{align*}
    T_{s_j} = \frac{1}{2} \left(\bigotimes_{k=1}^M T_0^k + is_j \bigotimes_{k=1}^M T_1^k \right)
\end{align*}
which also satisfies (\ref{eq:mirrormap_sym}). Since the initial state is also $\ket{\phi_0} = \ket{+}^{\otimes M}$, (\ref{eq:proof_expvalzero}) holds.

\section{Complex weights summation} \label{app:ComplexWeightProof}
From the definition (\ref{eq:complexweights_def}) we have
\begin{align}
    \sum_{\mathbf{S}} w_i^* (\mathbf{S}) w_j (\mathbf{S})  &=  \frac{1}{2^{2n}} \sum_{\mathbf{S}} \prod_{k=1}^n (-s_k i)^{i_k} (s_k i)^{j_k} \nonumber \\ &= \frac{1}{2^{2n}}  \prod_{k=1}^n \sum_{s_k=\pm 1} (-s_k i)^{i_k} (s_k i)^{j_k}
\end{align}
if $j \ne i$ then there is at least one factor for which $i_k = 0$ and $j_k=1$. For this factor
\begin{align}
    \sum_{\mathbf{S}} w_i^* (\mathbf{S}) w_j (\mathbf{S})   = \frac{1}{2^{2n}}   \sum_{s_k=\pm 1} s_k i = 0
\end{align}
making all the product zero. For $i=j$ we have
\begin{align}
    \sum_{\mathbf{S}} w_i^* (\mathbf{S}) w_i (\mathbf{S})  = \sum_{\mathbf{S}} \abs{w_i(\mathbf{S})}^2  =\sum_{\mathbf{S}} \frac{1}{2^{2n}} =\frac{1}{2^{n}} \, .
\end{align}

\section{Calculation of the average QFI}\label{app:AvgQFI}
\subsection{Symmetric coupling}
This section concerns the calculation of eq. (\ref{eq:Avgqfi}). Since $\expval{J_z} = 0$ (Appendix \ref{app:vanishingJz}) for all outcome trajectories at any stage of the protocol, the QFI in eq. (\ref{eq:FQ_def}) becomes
\begin{align}\label{eq:appendixQFI}
  F_Q\left[\ket{\phi_n (\mathbf{S})} ; J_z \right] &= 4 \expval{J_z^2} \nonumber \\ &=\frac{4}{P(\mathbf{S})} \Tr \ketbra{\phi_n(\mathbf{S})}{\phi_n(\mathbf{S})} J_z^2. 
\end{align}
where the second equality comes from the normalization condition $\braket{\phi_n (\mathbf{S})}{\phi_n (\mathbf{S})}=P(\mathbf{S})$.
Let us define the normalized \textit{average state}
\begin{align}
\label{eq:average_state}
    \overline{\rho}_n \equiv \sum_{\textbf{S}} \ketbra{ \phi_n(\mathbf{S})}{\phi_n(\mathbf{S})}. 
\end{align}
We emphasize the introduction of Eq. \eqref{eq:average_state} is only for the calculation of the averaged QFI Eq. \eqref{eq:avgfqproof1} below. In particular, the QFI of Eq. \eqref{eq:average_state} is not an appropriate precision bound of our sensing scenario, as the input probe state of each sensing interrogation is a random multi-cat state (the pure-state unravelling) $\ket{\phi_n({\textbf{S}})}$.

Then, we can write the average QFI as proportional to the expectation value of $J_z^2$ in $\overline{\rho}_n$
\begin{align}\label{eq:avgfqproof1}
    \overline{F}_{Q}^{(n)}= \sum_{\mathbf{S}} P(\mathbf{S}) F_Q\left[\ket{\phi_n (\mathbf{S})} ; J_z\right]= 4 \Tr  \overline{\rho}_n J_z^2.
\end{align}
Moreover, eq. (\ref{eq:symstate_kraus}) yields
\begin{align}
    \overline{\rho}_0 &= \ketbra{+}{+}^{\otimes M}\, , \nonumber\\ 
    \overline{\rho}_n &= \frac{1}{2} \left[T_0^{\otimes M} \overline{\rho}_{n-1}  T_0^{\otimes M \dagger} + T_1^{\otimes M} \overline{\rho}_{n-1}  T_1^{\otimes M \dagger} \right]\, , \label{eq:sym_kraus}
\end{align}
which means that each stage of the protocol acts on $\overline{\rho}_n$ as a unital quantum channel with Kraus operators $\frac{1}{\sqrt{2}} T_0^{\otimes M}$ and $\frac{1}{\sqrt{2}} T_1^{\otimes M}$. For unital channels with a given Kraus representation, the fixed points are those states that commute with each Kraus operator. Except in special cases (e.g. $\phi=\frac{\pi}{2}, \alpha=\frac{\pi}{4}$), the two Kraus operators have no eigenvectors in common, which means that the only state that commutes with both matrices is the maximally mixed state $\rho = \frac{1}{M+1} I_{M+1} $, hence this is the channel’s unique fixed point. Note that the 
the state belongs to the \textit{fully symmetric subspace} of fixed angular momentum $\frac{M}{2}$, which results in steady state being the identity in that subspace. This is a relevant remark for comparison with the non-symmetric scenario.
When $n \rightarrow \infty$, we can insert the fixed point of $\overline{\rho}_n$, i.e.  $\overline{\rho}_n \rightarrow \frac{1}{M+1} I_{M+1}$ into (\ref{eq:avgfqproof1}) giving eq. (\ref{eq:QFIsym_longtime}) in the main text. 
To obtain $\overline{F}_{Q}^{(n)}$ at a finite $n$, let us define the \textit{two-particle reduced state} as
\begin{align}
    \varrho_n (\mathbf{S}) &:=  \Tr_{M-2} \ketbra{\phi (\mathbf{S})}{\phi (\mathbf{S})} \nonumber \\ &= \sum_{pq=1}^{2^n}  w_p(\mathbf{S}) w_q^*(\mathbf{S})  \ketbra{\psi_p}{\psi_q}^{\otimes 2} \braket{\psi_p}{\psi_q}^{M-2},
\end{align}
where we used eq. (\ref{eq:sym_state}). Define now the \textit{average two-particle reduce state}
\begin{align}\label{eq:aveeffst_nonsym}
   \overline{\varrho}_n = \sum_{\mathbf{S}}  \varrho_n (\mathbf{S})
\end{align}
where $P(\mathbf{S})$ is cancelled out by the state normalization. Now we use the fact that 
\begin{align} \label{eq:complexWeights}
    \sum_{\mathbf{S}} w_p^*(\mathbf{S}) w_q(\mathbf{S}) = \frac{1}{2^n} \delta_{pq}
\end{align}
to get
\begin{align} \label{eq:rho2partavg_sym}
      \overline{\varrho}_n = \frac{1}{2^n} \sum_{p=1}^{2^n}  \ketbra{\psi_p}{\psi_p}^{\otimes 2}.
\end{align}
On the other hand, notice that $\overline{F}_{Q}^{(n)}$ is equal to
\begin{align}
\overline{F}_{Q}^{(n)} = 4  \sum_{\mathbf{S}} P(\mathbf{S}) \bra{\phi (\mathbf{S})} J_z^2 \ket{\phi (\mathbf{S})} \frac{1}{P(\mathbf{S})}\, ,
\end{align}
where we used eqs. (\ref{eq:Avgqfi}) and (\ref{eq:appendixQFI}). By inserting $J_z = \frac{1}{2} \sum_{m=1}^M \sigma_z^{m}$ into the expression above, we obtain
\begin{align}
\overline{F}_{Q}^{(n)} = M + M(M-1) \sum_{p,q=1}^{2^n} &\left(\sum_{\mathbf{S}}  w_p^*(\mathbf{S}) w_q(\mathbf{S}) \right) \nonumber \\ &\times \bra{\psi_p} \sigma_z \ket{\psi_q}^2 \braket{\psi_p}{\psi_q}^{M-2},
\end{align}
from eq. (\ref{eq:complexWeights}) we get
\begin{align}
\overline{F}_{Q}^{(n)} = M + M(M-1) \frac{1}{2^n}\sum_{p=1}^{2^n} \bra{\psi_p} \sigma_z \ket{\psi_p}^2 \, .
\end{align}
This yields eq. (\ref{eq:QFI_symmm}) in the main text
\begin{align} \label{eq:fqn}
\overline{F}_{Q}^{(n)} = M + M(M-1) \overline{H}_S(n)
\end{align}
with 
\begin{align}
  \overline{H}_S(n) := \Tr \overline{\varrho}_n \sigma_z^{\otimes 2}.
\end{align}
For finite $n$, we can compute $\overline{\varrho}_n $ numerically and thus $ \overline{H}_S(n)$ as follows
\begin{align*}
    \overline{\varrho}_0 &= \ketbra{+}{+}^{\otimes 2}\, , \\ 
    \overline{\varrho}_n&= \frac{1}{2} \bigg[ T_0^{\otimes 2}  \overline{\varrho}_{n-1}  T_0^{\otimes 2\dagger} + T_1^{\otimes 2} \overline{\varrho}_{n-1}  T_1^{\otimes 2\dagger}   \bigg]\, .
\end{align*}
As we have shown above, the average QFI for $n\rightarrow \infty$ is $\overline{F}_{Q}^{(\infty)} = \frac{M (M+2)}{3}$, which, together with Eq. (\ref{eq:fqn}) gives
\begin{align}
 \overline{H}_S(\infty) =  \frac{\overline{F}_{Q}^{(\infty)}-M}{M(M-1)}=\frac{1}{3}  
\end{align}

\subsection{Non-symmetric coupling}

We proceed in the same fashion as in the symmetric case. The average QFI is given by
\begin{align}
    F_Q \left[  \ket{\phi (\mathbf{S})} ; J_z \right] = 4  \expval{J_z^2} \, .
\end{align}
Using $ \ket{\phi (\mathbf{S})} = \sum_{i=1}^{2^n} w_i (\mathbf{S}) \bigotimes_{k=1}^M \ket{\psi_i^k}$ and (\ref{eq:complexWeights}), we have
\begin{align}
    \overline{F}_Q^{(n)} = \sum_{\mathbf{S}} P(\mathbf{S})  F_Q \left[  \ket{\phi (\mathbf{S})} ; J_z \right] =  M +  2 \overline{H}_{NS} (M, n)
\end{align}
where
\begin{align}\label{eq:h2_def_nonsym}
    \overline{H}_{NS}(M, n) := \frac{1}{2^n}\sum_{i<j}^M \sum_{p=1}^{2^n}  \bra{\psi_p^i} \sigma_z \ket{\psi_p^i} \bra{\psi_p^j} \sigma_z \ket{\psi_p^j}.
\end{align}
First, we obtain an asymptotic expression for $\overline{F}_Q^{(n)}$ when $n\rightarrow \infty$. As we know from the symmetric coupling scenario, $\expval{J_z}=0$ implies
\begin{align} \label{eq:FQznonsym_4}
    \overline{F}_Q^{(n)} = 4 \Tr \overline{\rho}_n J_z^2
\end{align}
where $\overline{\rho}$ is given by (\ref{eq:average_state}). Furthermore, we can express $\overline{\rho}$ via the Kraus map (\ref{eq:nonsym_kraus}) to get
\begin{align} 
    \overline{\rho}_n = \frac{1}{2} \left[ \bigotimes_{k=1}^M T_0^k  \overline{\rho}_{n-1}  \bigotimes_{k=1}^M T_0^{k \dagger}  +\bigotimes_{k=1}^M T_1^k  \overline{\rho}_{n-1}  \bigotimes_{k=1}^M T_1^{k\dagger}  \right]
\end{align}
which shows that $\overline{\rho}_n$ evolves by a unital channel with Kraus operators $\frac{1}{\sqrt{2}} \bigotimes_{k=1}^M T_0^k$, $\frac{1}{\sqrt{2}} \bigotimes_{k=1}^M T_1^k$. This time, the state does not have symmetry under particle exchange, and therefore it cannot be described only within the fully symmetric subspace. The channel's fixed point is then the identity in the whole $2^M$ dimensional Hilbert space $\rho = \frac{1}{2^M} I_{2^M}$. We obtain our result by replacing $\overline{\rho}_n \rightarrow \frac{1}{2^M} I_{2^M}$ in (\ref{eq:FQznonsym_4}), obtaining
\begin{align}
     \overline{F}_Q^{(\infty)} = 4 \frac{1}{2^M} \Tr J_z^2 = M.
\end{align}
For the finite $n$ expression the procedure is analogous to the symmetric coupling scenario, with slight changes due to the lack of symmetry under particle exchange. Let us define the \textit{two-particle reduced state} in the non-symmetric case as
\begin{align}
    \varrho_n^{k,m} (\mathbf{S}) :=  \Tr_{M-2 \ne k,m} \ketbra{\phi (\mathbf{S})}{\phi (\mathbf{S})}
\end{align}
meaning that we are tracing out all particles but $k, m$. The \textit{average} two-particle reduced state is
\begin{align}\label{eq:aveeffst_nonsym}
   \overline{\varrho}_n^{k,m} = \sum_{\mathbf{S}}  \varrho_n^{k,m} (\mathbf{S})
\end{align}
where $P(\mathbf{S})$ is cancelled out by the state normalization. We use (\ref{eq:complexWeights}) to get
\begin{align} \label{eq:avgrho_nonsym}
      \overline{\varrho}_n^{k,m} = \frac{1}{2^n} \sum_{p=1}^{2^n}  \ketbra{\psi_p^k}{\psi_p^k} \otimes \ketbra{\psi_p^m}{\psi_p^m} 
\end{align}
which allows us to write (\ref{eq:h2_def_nonsym}) as
\begin{align}\label{eq:hNS_nonsym}
     \overline{H}_{NS}(M, n) = \sum_{k<m}^M \Tr \overline{\varrho}_n^{k,m} \sigma_z^{\otimes 2}
\end{align}
and therefore
\begin{align} 
    \overline{F}_{Q}^{(n)}  = M + 2\overline{H}_{NS}(M, n).
\end{align}
 The average two-particle reduced density matrix can be efficiently computed numerically by the following formulas
\begin{align} \label{eq:numQFI_nonsym}
    \overline{\varrho}_0^{k,m} &= \ketbra{+}{+}^{\otimes 2}\, , \nonumber \\ 
    \overline{\varrho}_n^{k,m} &= \frac{1}{2} \bigg[ \left(T_0^k \otimes T_0^m \right) \overline{\varrho}_{n-1}^{k,m}  \left(T_0^k \otimes T_0^m \right)^\dagger \nonumber \\ &+ \left(T_1^k \otimes T_1^m \right) \overline{\varrho}_{n-1}^{k,m}  \left(T_1^k \otimes T_1^m \right)^\dagger    \bigg]
\end{align}
which follow directly from (\ref{eq:state2}).

\section{$\overline{H}_S$ and $\overline{H}_{NS}$  } \label{app: H2avg}
\subsection{Symmetric coupling}
From (\ref{eq:rho2partavg_sym}) we have
\begin{align}
    \overline{\varrho}_n = \frac{1}{2^n}  \sum_{i=1}^{2^n} \ketbra{\psi_i}{\psi_i}^{\otimes 2},
\end{align}
by inserting
\begin{align}
   \ket{\psi_i} = a_i \ket{0} + b_i \ket{1}.
\end{align}
we get
\begin{align}
 \overline{\varrho}_n = \frac{1}{2^n} \sum_{i=1}^{2^n} \left(a_i \ket{0} + b_i \ket{1} \right)^{\otimes 2} \left(a^*_i \bra{0} + b^*_i \bra{1} \right)^{\otimes 2}
\end{align}
which yields
\begin{align}
    \overline{H}_S (n) &= \Tr{\sigma_z^{\otimes 2} \overline{\varrho}_n} \nonumber \\ &= \frac{1}{2^n} \sum_{i=1}^{2^n} \left(\abs{a_i}^2 - \abs{b_i}^2 \right)^2.
\end{align}
Now let us write down the coefficients with the usual parametrization $a_i = \cos \left(\frac{\theta_i}{2} \right)$, $b_i = \text{e}^{i\varphi_i} \sin \left(\frac{\theta_i}{2} \right)$, where $(\theta_i, \varphi_i)$ are the coordinate angles of $ \ket{\psi_i}$ in the Bloch sphere representation. This yields
\begin{align}
    \overline{H}_S(n) = \frac{1}{2^n} \sum_{i=1}^{2^n} \cos^2 \left(\theta_i \right) \leq 1
\end{align}
which only vanishes if all $\theta_i = \frac{(2m+1)\pi }{2}$. With this we have finalized to prove that $\overline{H}_S(n)$ is a positive function, that cannot decrease with $n$ and does not vanish.

\subsection{Non-symmetric coupling}
Inserting $\ket{\psi_i^k} = a_{ik} \ket{0} + b_{ik} \ket{1}$ in (\ref{eq:avgrho_nonsym}) we get
\begin{align}
    \overline{\varrho}^{k,m}_n  &=\nonumber \\ &\frac{1}{2^n}  \sum_{i=1}^{2^n} \left(a_{ik} \ket{0} + b_{ik} \ket{1} \right) \left(a^*_{ik} \bra{0} + b^*_{ik} \bra{1} \right)\nonumber \\ &\otimes \left(a_{im} \ket{0} + b_{im} \ket{1} \right) \left(a^*_{im} \bra{0} + b^*_{im} \bra{1} \right),
\end{align}
carrying on an analogous procedure as in the symmetric case, we obtain
\begin{align}
     \overline{H}_{NS}(M, n) &= \sum_{k<m}^M \Tr \overline{\varrho}_n^{k,m} \sigma_z^{\otimes 2} \nonumber \\ &= \frac{1}{2^n} \sum_{k<m}^M  \sum_{i=1}^{2^n} \cos \left(\theta_{ik} \right) \cos \left(\theta_{im} \right) 
\end{align}
where we replaced $a_{ik} = \cos \left(\frac{\theta_{ik}}{2} \right)$, $b_{ik} = \text{e}^{i\varphi_{ik}} \sin \left(\frac{\theta_{ik}}{2} \right)$. Note that, since for a fixed $k,m$ we have $\theta_{im} \ne \theta_{ik}$, the terms can sum up destructively. In general 
\begin{align}
     \overline{H}_{NS}(M, n) < \frac{M(M-1)}{2}\, .
\end{align}

\section{Master equation description} \label{app:MEderivation}
In this appendix we detail the derivation of the LMEs (\ref{eq:master_eq_sym}) and (\ref{eq:MEdisorder}) for the unconditional evolution of the spin ensemble in
the limit \(dt, \phi ,\alpha \ll 1\). Recall that the spin ensemble evolves, in the general (non-symmetric coupling) case, according to the sequential
map (\ref{eq:state2}) conditioned on the result \(\left| \pm _y\right\rangle\) of the \(n\)-th measurement of the central spin, where the one-spin operators are
defined in Eqs. (\ref{eq:T^k_nonsymmetric}). In the limit \(\phi ,\alpha \ll 1\), we can expand Eqs. (\ref{eq:T^k_nonsymmetric}) to the first order in $dt$. Correspondingly, the normalized average state,
\(\bar{\rho} (t)=\bar{\rho} _n\), cf. Eq. (\ref{eq:average_state}), evolves according to
\begin{equation}
\label{eq:ME_general}
\frac{d\bar{\rho} }{dt}=-i\left[\dot{\phi }J_z,\bar{\rho} \right]+4\frac{1}{dt}\mathcal{D}\left[\sum _i\alpha _i\sigma _x^i\right]\bar{\rho},
\end{equation}
with $\dot{\phi}\equiv \phi/dt$.
In particular, for the symmetric-coupling { }case, \(\alpha _i=\alpha\), this reduces to Eq. (\ref{eq:master_eq_sym}) of the main text.

Now consider the nonsymmetric-coupling case, where we assume \(\alpha _i=\alpha +\delta _i\), with \(\alpha\) the mean coupling strength and \(\delta
_i\) the disorder associated with the \(i\)-th spin. Eq. (\ref{eq:ME_general}) thus repesents the unconditional evolution of the spin ensemble for a specific realization
of the disorder \(\left\{\delta _i\right\}\). Denoting \(\rho ^{\rm S}(t)\) as the solution for the symmetric case, we have the expansion
\begin{equation}
\label{eq: me_disorder_G}
\rho ^{\text{NS}}(t)=\rho ^{\rm S}(t)+\delta  C_1(t)+\delta ^2C_2(t)+\text{...}
\end{equation}
where \(\delta =\sqrt{\overline{\delta _i^2}}\) characterizes the strength of the disorder. By expanding the RHS of Eq. (\ref{eq:ME_general}) in orders \(\delta\), we
arrive at series of differential equations for the evolution of \(C_1(t), C_2(t),\dots\) In particular, first-order expansion in \(\delta\) gives
\begin{align}
\frac{d}{dt}C_1(t) = &-i\left[\dot{\phi }J_z,C_1(t)\right]+\frac{4\alpha}{\delta  dt} \Big[  J_x\rho ^{\rm S}(t)\mu\nonumber\\
&-\frac{1
}{2} \left\{J_x,\mu\right\}\rho ^{\rm S}(t)-\frac{1 }{2}\rho^{\rm S}(t) \left\{J_x,\mu\right\}\Big],
\end{align}
where we have defined $\mu=\sum _i\delta _i\sigma _x^i$. Averaging over the disorder, we find
\begin{equation}
\frac{d}{dt}\overline{C_1(t)}=-i\left[\dot{\phi }J_z,\overline{ C_1(t)}\right]\, .
\end{equation}
Since we initialize the spin ensemble in a fully symmetric state $\ket{+}^{\otimes M}$, we have \(  \overline{C_1(0)}=0\) which leads to \( \overline{C_1(t)}=0\). Therefore, by averaging Eq.~\eqref{eq: me_disorder_G} with respect to disorder realizations, we have
\begin{equation}
\bar{\rho} ^{\text{NS}}(t)=\rho ^{\rm S}(t)+\delta ^2\overline{C_2(t)}+\text{...} \, .
\end{equation}
Similarly, second-order expansion of Eq.~\eqref{eq:ME_general} in \(\delta\) gives
\begin{align}
\label{eq: NSrho_traj}
\frac{d}{dt}C_2(t)=-i\left[\dot{\phi }J_z,C_2(t)\right]+\frac{4}{\delta ^2dt}\mathcal{D}\left[\mu\right]\rho ^{\rm S}(t)\nonumber\\
+\frac{4\alpha
^2}{dt}\mathcal{D}\left[J_x\right]C_2(t)
+\frac{4\alpha}{\delta ^2dt} \Big[  J_xC_1(t)\mu\nonumber\\
-\frac{1}{2} \left\{J_x,\mu\right\}C_1(t)-\frac{1}{2}C_1(t) \left\{J_x,\mu\right\}\Big].
\end{align}
Now we average Eq. \eqref{eq: NSrho_traj} with respect to the disorder. To arrive at a closed equation, we adopt the approximation $\overline{C_1\mu}\simeq \overline{C_1}\overline{\mu}$ which is an excellent approximation for weak disorder. We thus arrive at
\begin{align}
\label{eq: NSrho_average}
\frac{d}{dt}\overline{C_2(t)}=&-i\left[\dot{\phi }J_z,\overline{C_2(t)}\right]+\frac{4}{\delta ^2dt}\sum _i\overline{\delta _i^2}\mathcal{D}\left[\sigma
_x^i\right]\rho ^{\rm S}(t)\nonumber\\
&+\frac{4\alpha ^2}{dt}\mathcal{D}\left[J_x\right]\overline{C_2(t)}\, .
\end{align}
Putting together Eq. \eqref{eq:master_eq_sym} and Eq. \eqref{eq: NSrho_average}, we find
\begin{align}
\frac{d\bar{\rho} ^{\text{NS}}(t)}{dt}=&-i\left[\dot{\phi }J_z,\bar{\rho} ^{\text{NS}}\right]+\frac{4\alpha ^2}{dt}\mathcal{D}\left[J_x\right]\bar{\rho}
^{\text{NS}}\nonumber\\
&+\frac{4}{dt}\sum _{ i}\overline{\delta _i^2}\mathcal{D}\left[\sigma _i^x\right]\bar{\rho}^{\text{NS}},
\end{align}
which recovers Eq. \eqref{eq:MEdisorder}.
\bibliography{bib}
\bibliographystyle{quantum}

\end{document}